\documentclass[10pt,pra,superscriptaddress,twocolumn,nofootinbib]{revtex4-2}
\usepackage[utf8]{inputenc}
\usepackage{CJKutf8}
\usepackage{multirow}
\usepackage{amsmath,amssymb,amsthm}
\usepackage{times,ragged2e}
\usepackage[dvipsnames]{xcolor}
\usepackage[T1]{fontenc}

\usepackage{tikz}
\usetikzlibrary{quantikz2}
\usepackage{braket}
\usepackage{mathtools}
\usepackage{bbold}
\usepackage[hyperindex,breaklinks,colorlinks=true,citecolor=blue,urlcolor=blue]{hyperref}
\usepackage{hhline}
\usepackage{float,placeins}

\newcommand{\NS}{\mathcal{N\!S}}
\newcommand{\NSnotAtoB}{\mathcal{N\!S}_{A\not\to B}}
\newcommand{\NSnotBtoA}{\mathcal{N\!S}_{B\not\to A}}
\newcommand{\Q}{\mathcal{Q}}
\renewcommand{\L}{\mathcal{L}}
\renewcommand{\H}{\mathcal{H}}

\newcommand{\vecf}{\vec{f}}
\newcommand{\vecP}{\vec{P}}
\newcommand{\vecPstar}{\vec{P}^\mathrm{NS}_\star}

\newcommand{\iid}{{\em i.i.d.}}

\newcommand{\tr}{{\rm tr}}
\newcommand{\DKL}{D_\mathrm{KL}}
\newcommand{\CirL}{\mathcal{C}_\mathrm{L}}
\newcommand{\CirNL}{\mathcal{C}_\mathrm{NL}}

\DeclareMathOperator*{\argmin}{argmin}

\usepackage[capitalize]{cleveref}

\begin{document}

\begin{CJK*}{UTF8}{bsmi}

\title{Almost device-independent calibration beyond Born's rule:\\ Bell tests for cross-talk detection}

\author{Gelo Noel M. Tabia}
\email{gelo.tabia@foxconn.com}
\thanks{Corresponding author}
\affiliation{Hon Hai (Foxconn) Research Institute, Taipei, Taiwan}
\affiliation{Department of Physics and Center for Quantum Frontiers of Research \& Technology (QFort), National Cheng Kung University, Tainan 701, Taiwan}
\affiliation{Physics Division, National Center for Theoretical Sciences, Taipei 106319, Taiwan}

\author{Alex Yueh-Ting Shih}
\affiliation{Department of Physics and Center for Quantum Frontiers of Research \& Technology (QFort), National Cheng Kung University, Tainan 701, Taiwan}

\author{Jin-Yuan Zheng}
\affiliation{Department of Physics and Center for Quantum Frontiers of Research \& Technology (QFort), National Cheng Kung University, Tainan 701, Taiwan}

\author{Yeong-Cherng Liang}
\email{ycliang@mail.ncku.edu.tw}
\thanks{Corresponding author}
\affiliation{Department of Physics and Center for Quantum Frontiers of Research \& Technology (QFort), National Cheng Kung University, Tainan 701, Taiwan}
\affiliation{Physics Division, National Center for Theoretical Sciences, Taipei 106319, Taiwan}
\affiliation{Perimeter Institute for Theoretical Physics, Waterloo, Ontario, Canada, N2L 2Y5}

\begin{abstract}
In quantum information, device-independent protocols offer a new approach to information processing tasks, making minimal assumptions about the devices used. Typically, since these protocols draw conclusions directly from the data collected in a meaningful Bell test, the no-signaling conditions, and often even Born's rule for local measurements, are taken as premises of the protocol. Here, we demonstrate how to test such premises in an (almost) device-independent setting, i.e., directly from the raw data and with minimal assumptions.
In particular, for IBM's quantum computing cloud services, we implement the prediction-based ratio protocol to characterize how well the qubits can be accessed locally and independently.  More precisely, by performing a variety of Clauser-Horne-Shimony-Holt-type experiments on these systems and carrying out rigorous hypothesis tests on the collected data, we provide compelling evidence showing that some of these qubits suffer from measurement cross-talks, i.e., their measurement statistics are affected by the choice of measurement bases on another qubit.
Unlike standard randomized benchmarking, our approach does not rely on assumptions such as gate-independent Markovian noise. Moreover, despite the relatively small number of experimental trials, the direction of ``signaling'' may also be identified in some cases. Our approach thus serves as a complementary tool for benchmarking the local addressability of quantum computing devices.
\end{abstract}
\date{\today}
\maketitle

\section{Introduction}

The device-independent (DI)~\cite{Brunner2014:RMP} approach to physics can be traced back to Bell~\cite{Bell64} when he proved that local-hidden-variable (LHV) theories necessarily fail to reproduce some predictions of quantum theory. His proof relies only on the correlations among measurement outcomes conditioned on the chosen measurement settings. Thus, it requires no further knowledge about how the devices function. Since then, a few other no-go theorems based on the violation of Bell-like inequalities have also been obtained  (see, e.g.,~\cite{BPA+12,PRB+14,BUG+20}).

Apart from quantum foundational issues, the DI methodology also finds applications in several cryptographic tasks, such as randomness expansion~\cite{Colbeck09,Pironio10,Colbeck11} and key distribution~\cite{Barrett05,Acin07,Vazirani14}. In these DI protocols, it is crucial that the correlations obtained from the Bell experiment satisfy the so-called no-signaling~\cite{Popescu:FP:1994} (NS) conditions. Often, the security analysis further assumes that quantum theory is correct, in particular, that the outcome probabilities are specified by Born's rule for local measurements (see~\cite{Zapatero:2023aa,Primaatmaja2023} for a recent review).

In this work, we focus on applications of the DI approach to the characterization of quantum devices (see, e.g.,~\cite{Brunner08,Bancal11,Moroder13,LRB+15,SLChen16, SLChen18, Bancal:PRL:2018,QBW+19,Wagner2020,CMBC21,Sekatski2018}).
One of the requirements for the proper functioning of quantum computers is the ability to protect fragile quantum states from noise~\cite{DiVincenzo:2000aa}.
However, in some quantum computers, due to the proximity of the qubits and their high level of interconnectivity, it is conceivable that the interaction with a targeted qubit could simultaneously affect the state of the neighboring qubits.
To correct the errors from such cross-talk~\cite{Sarovar2020detectingcrosstalk} and other unwanted effects, we need some way to identify and quantify the noise in a quantum device. The most widely used approaches for this task are based on randomized benchmarking (RB)~\cite{Emerson_2005-RB,Levi2007-RB,Knill2008-RB,Magesan2011-RB} or
gate-set tomography (GST)~\cite{Blume-Kohout2017-GST,Nielsen2021-GST-Quantum} (see also~\cite{MGS+13}).

In a typical RB method, we measure the error rate of a particular set of quantum gates by applying a sequence of random gates that would ideally correspond to an identity operation if the gates were perfect. Meanwhile, GST is a method that incorporates elements of quantum process tomography into a procedure that also deals directly with state preparation and measurement (SPAM) errors. GST inherits some of the problems of tomographic methods, particularly the need for large samples to estimate noise parameters.
To achieve sample efficiency, one can turn a GST protocol into a randomized scheme and use classical shadow estimation techniques~\cite{Huang2020-CSE,Paini2021-CSE} that allow one to deduce various linear functionals of the gate-set noise~\cite{Helsen2023-SEGS}.

However, both RB and GST often involve the assumption of temporally uncorrelated noise.
In RB, the exponential decay in the average gate-sequence fidelity assumes that the noise is Markovian, and one can even identify the presence of non-Markovian noise by the failure of the exponential model~\cite{Epstein2014-RB,Fogarty2015-RB,Wallman2018randomized}.
Similarly, in GST, a Markovian noise model is used so that the contributions of SPAM and gate-set errors can be estimated separately. While there have been recent attempts to incorporate non-Markovian noise~\cite{FigueroaRomero2021-RB, FigueroaRomero2022towardsgeneral}, it is natural to wonder whether one can identify unwanted cross-talks using only {\em minimal assumptions}.

Here, building on earlier studies~\cite{LZ19}, we show that it is indeed possible to certify---in an {\em almost} DI manner---the presence of cross-talks directly from the raw measurement data obtained from a quantum computer. By ``almost'' DI, we mean that we consider standard DI assumptions on our Bell tests but with the usual assumption of measurement independence~\cite{PRB+14} (more commonly known as the ``freedom of choice" assumption) replaced by the {\em assumption} that the pseudo-random string of inputs does {\em not} alter the behavior of the individual qubits, even though we generate the inputs and feed them to the device preparing the qubits {\em before} their preparation.

Importantly, experimental trials are not necessarily independent and identically distributed ({\em i.i.d.}). In particular, assuming that the trials are \iid~when they are not may open the so-called memory loophole~\cite{BCH+02}. Even if the trials are {\em i.i.d.}, statistical fluctuations may still render the relative frequencies of the measurement outcomes---taken as a proxy of the underlying correlation---incompatible with the NS constraints. To cope with this complication in the context of DI certification, various methods for regularizing the relative frequencies to the set of correlations compatible with the NS constraints~\cite{Bancal:2014aa,Bernhard:2014aa} (or even outer approximations of the quantum set~\cite{Schwarz16,Lin_PointEstimation_2018}) have been proposed. However, one can also adopt a more rigorous approach based on hypothesis testing.

Indeed, in statistical inference, it is customary to report the $p$-value for a null hypothesis to be correct. Here, we follow~\cite{LZ19,CCC+24,Patra24} and consider the prediction-based-ratio (PBR) protocol~\cite{ZGK11,ZGK13} for upper bounding the $p$-value on the plausibility of a given null hypothesis in producing the data observed in a Bell test. The PBR
protocol was originally introduced as a rigorous statistical tool for rejecting the null hypothesis associated with LHV theories. In~\cite{CCC+24}, it was adapted to perform DI certification of desirable quantum properties (e.g., those discussed in~\cite{Brunner08,Moroder13,LRB+15,Bancal15}) with a confidence interval. Notice that these certification tasks presuppose Born's rule for local measurements and, hence, compatibility with the NS constraints. In this work, we illustrate how the PBR protocol can be used to reveal a violation of these premises and, consequently, the presence of cross-talks in real quantum devices with a relatively small sample size.

We structure the rest of this paper as follows. \cref{Sec:Prelim} introduces our notations and recalls the background knowledge required for analyzing the data collected in a Bell test. Then, in~\cref{App:IBM}, we explain how we apply the PBR protocol to the data collected from ``Bell tests'' performed on IBM Quantum (IBMQ) devices. We then present our results in~\cref{Sec:Results} and end with further discussions in~\cref{Sec:Discussion}.

\section{Preliminaries}~\label{Sec:Prelim}

\subsection{No-signaling conditions and the no-signaling set}\label{Sec:Sets}

For simplicity, we consider only the simplest, bipartite Bell scenario where two parties, Alice and Bob, with two inputs and two outputs each. If we denote Alice's (Bob's) inputs/ settings by $x\in X$ ($y\in Y$) and outputs/ outcomes by $a\in A$ ($b\in B$), then a Bell correlation $\vec{P}:= \{P(a,b|x,y)\}$ is the collection of joint conditional probability distributions of measurement outcomes given the choice of settings.

If we require that Bob cannot signal his input choice ($y$ or $y'$) to Alice, then her marginal probabilities must satisfy
\begin{subequations}\label{Eq:NS}
\begin{align}
\label{eq:Alice_OWNS}
    P(a|x) = \sum_{b} P(a,b|x,y) = \sum_{b} P(a,b|x,y'),
    \forall\,\, a,x,y,y'.
\end{align}
In this case, we say that $\vec{P}$ is one-way no-signaling (OWNS) from Bob to Alice, and we denote the set of all such correlations by $\NSnotBtoA$.
On the other hand, if, instead, we require that Alice cannot signal her input choice ($x$ or $x'$) to Bob, then we have
\begin{align}
\label{eq:Bob_OWNS}
    P(b|y) = \sum_{a} P(a,b|x,y) = \sum_{a} P(a,b|x',y),
    \forall\,\, b,y,x,x'.
\end{align}
\end{subequations}
We refer to $\vec{P}$ satisfying~\cref{eq:Bob_OWNS} as being OWNS from Alice to Bob, and we denote the set of such correlations accordingly by $\NSnotAtoB$.

The set $\NS$ of (two-way) no-signaling (NS) correlations, defined by~\cref{Eq:NS}, is the intersection of the two OWNS sets $\NSnotAtoB$ and $\NSnotBtoA$.
Originally, the NS conditions of~\cref{Eq:NS} were inspired by the notion of relativistic causality from special relativity~\cite{Popescu:FP:1994}, which prohibits a causal influence between spacelike separated parties.
In our work, we provide an alternative interpretation of the NS conditions in the context of measurement cross-talk effects: if there is no unintended measurement cross-talk between the qubits,  the choice of measurement basis on one qubit will have no impact on the marginal measurement statistics of any other qubit. In this case, the NS conditions of \cref{Eq:NS} follow. In other words, the violation of any constraint from \cref{Eq:NS} is a signature of measurement cross-talks, modulo the assumption mentioned in the Introduction.

\begin{figure}
    \centering    \includegraphics[width=0.95\columnwidth]{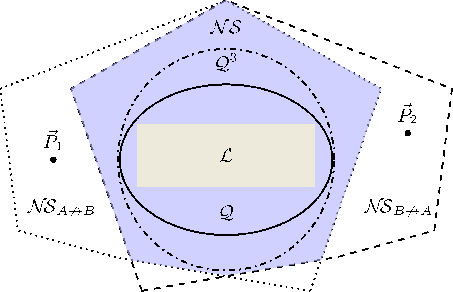}
\caption{Schematic illustrating the inclusion relations of \cref{Eq:Inclusions}.  A correlation or relative frequency such as $\vecP_1$ lies outside $\NSnotBtoA$ (dashed-boundary polygon), and, hence, also outside $\NS$ (filled polygon), $\Q^3$ (dashed-dotted ellipse), $\Q$ (solid ellipse), and $\L$ (shaded rectangle). Similarly, a correlation or relative frequency such as $\vecP_2$ lies outside $\NSnotAtoB$ (dotted-boundary polygon), and, hence, also outside $\NS$, $\Q^3$,  $\Q$, and $\L$. Note that the Figure is not meant to capture {\em all} geometrical aspects of these sets. For example, $\L$ and $\NS$, and hence all sets sandwiched between them, cf.~\cref{Eq:Inclusions}, share common boundaries~\cite{Brunner2014:RMP}, but this and various other geometrical features~\cite{Goh2018,CTC+23} are not depicted in the Figure.}
\label{Fig:Inclusions}
\end{figure}

In a Bell test, we are also often interested in two particular subsets of $\NS$: the set $\L$ of (Bell-)local~\cite{Brunner2014:RMP} correlations and the set $\Q$ of quantum correlations.
We have $\vecP \in \L$ if there exists an LHV $\lambda$ satisyfing a normalized distribution $p(\lambda)\ge 0$ and local deterministic response functions $P_A(a|x,\lambda), P_B(b|y,\lambda)=0,1$ with $\sum_a P_A(a|x, \lambda) = 1 = \sum_b P_B(b|y,\lambda)$ such that for all $a,b,x,y$, we can write~\cite{Bell64,Brunner2014:RMP}
\begin{equation}
\label{eq:localP}
    P(a,b|x,y)\overset{\L}{=}\sum_\lambda p(\lambda) P_A(a|x,\lambda) P_B(b|y,\lambda).
\end{equation}
Otherwise, we say that $\vecP$ is (Bell-)nonlocal.
Meanwhile, we have $\vecP \in \Q$ if it can be obtained from local measurements performed by Alice and Bob on a shared quantum state $\rho_{AB}$, then Born's rule dictates that
\begin{equation}
    \label{eq:quantumP}
    P(a,b|x,y) \overset{\Q}{=} \tr(\rho_{AB} M^A_{a|x}\otimes M^B_{b|y}),
\end{equation}
where $M^A_{a|x} (M^B_{b|y})$ denotes the positive operator-valued measure element associated with outcome $a$ ($b$) of Alice's (Bob's) $x$-th ($y$-th) measurement setting.
It is easy to verify that $\L$, $\Q$, and $\NS$ are convex sets that satisfy the strict inclusion
\begin{equation}\label{Eq:Inclusions}
	\L \subset \Q \subset \Q^k\subset \NS=\NSnotBtoA \cap \NSnotAtoB\quad\forall\,\,k,
\end{equation}
where $\Q^k$ is an outer NS approximation of $\Q$ (more on this below). See \cref{Fig:Inclusions} for a diagrammatic representation.

When the cardinalities of $X, Y, A$, and $B$ are finite, $\L$ is a convex polytope, i.e., the convex hull of a finite set of extreme points.  In contrast, since the quantum set $\Q$ is not a polytope, there is generally no simple criterion to test whether a correlation $\vecP$ belongs in $\Q$. Nevertheless, various outer approximations of $\Q$ (see, e.g.,~\cite{NPA,NPA2008,Doherty08,Moroder13}) facilitate its membership test via a sequence of supersets $\Q^k$ such that $\NS\supset \Q^1 \supset \Q^2 \supset \cdots \supset \Q$. In the following, we use the level-3 of the Moroder hierarchy~\cite{Moroder13}, denoted by $\Q^3$, as our outer approximation of $\Q$. This choice is motivated by the observation in~\cite{Lin:Quantum:2022} (see, e.g., Table 2 and the top left subplot of Fig. 3 therein) that the lowest level of the hierarchy from either~\cite{Moroder13,NPA} is visibly not tight, but going to a level even higher than $\Q^3$ may not worth the extra computation time. However, from the analysis of~\cite{Lin:Quantum:2022}, we expect similar results to hold if we adopt other outer approximations with similar computational complexity.

\subsection{Hypothesis testing and the prediction-based-ratio method}\label{Sec:PBR}

Often, we perform an experiment to test a particular (null) hypothesis, such as that derived from a theoretical prediction.
In statistical hypothesis testing, one effective way of determining the plausibility of a null hypothesis $\H$ from experimental data is to compute a $p$-value upper bound from some real function of the data called a test statistic $T$.
The $p$-value then represents the tail probability for the observed value of $T$ conditioned on $\H$, i.e., if the observed value of $T$ is $t$, then
\begin{equation}
	p\text{-value} =\text{Prob}(T\ge t| \H \text{ holds}),
\end{equation}
which tells us how likely the data can be explained by the hypothesis $\H$.

Historically, Bell tests were introduced to determine if Nature is compatible with the description of LHV theories.
However, any real Bell test necessarily involves only a finite number of trials where we obtain the counts of events involving different combinations of inputs and outputs.
To cope with this limitation, the prediction-based-ratio (PBR) protocol---motivated by an earlier work of Gill~\cite{Gill03}---was introduced to provide a systematic, efficient method for upper bounding the corresponding $p$-value. In~\cite{LZ19}, it was noted that the PBR protocol can be straightforwardly adapted to test the plausibility of other physical theories, including a general NS theory.

For concreteness, suppose we conduct a Bell test with a total of $N$ trials. In each trial, the inputs $x$ and $y$ are chosen randomly according to some fixed distribution $P(x,y)$.
Thus, the data generated in each trial is a set of four numbers $(a,b,x,y)$.
For definiteness, consider now the hypothesis that the data observed is generated by an underlying NS process describable by some correlation $\vecP \in \NS$, which may vary from one trial to the next.

Even if the experimental trials are {\em i.i.d.}, the data alone will not allow us to identify $\vecP$ exactly.
Nonetheless, we can estimate $ \vecP$ by computing the relative frequencies $\vec{f}:=\{f(a,b|x,y)\}$ for each outcome pair $(a,b)$ given the choice of input pair $(x,y)$,
\begin{equation}\label{Eq:f}
    f(a,b|x,y):= N_{a,b,x,y}/N_{x,y},
\end{equation}
where $N_{a,b,x,y}$ is the number of trials where the input-output combination $(a,b,x,y)$ occurs,  $N_{x,y}:= \sum_{a,b} N_{a,b,x,y}$, and
\begin{equation}\label{Eq:SumNxy}
	\sum_{x,y} N_{x,y} = N.
\end{equation}
In the asymptotic limit where $N\to\infty$, statistical fluctuations vanish, and therefore $\vec{f}$ approaches $\vec{P}$.
For (finite) \iid~trials, the amount of statistical evidence in the data contrary to our hypothesis can be measured~\cite{vanDam2005o} in terms of the Kullback-Leibler (KL) divergence.

More precisely, if we believe the NS hypothesis to be true, the ``best-fitting'' NS correlation would be given by the minimizer of the following optimization problem:
\begin{align}
    \DKL (\vec{f}||\NS) &= \min_{\vecP \in \NS} \sum_{a,b,x,y} P(x,y) f(a,b|x,y) \nonumber \\
    & \qquad \times \log\left[ \frac{f(a,b|x,y)}{P(a,b|x,y)}\right]\label{Eq:DKL}.
\end{align}
Importantly, this optimization can be efficiently solved using a numerical solver such as MOSEK~\cite{mosek}. In~\cite{githubCode}, we provide an implementation of this optimization in MATLAB via YALMIP~\cite{YALMIP}.\footnote{For the results presented in~\cref{Sec:Results}, we also use a somewhat more accurate implementation of \cref{Eq:DKL} via PENLAB~\cite{PENLAB} (courtesy of Denis Rosset), which generally gives a tighter $p$-value upper bound.}
Since $\NS$ is a convex set and the KL divergence is a {\em strictly} convex function of $\vecP$, the minimizer $\vecPstar$ of the above optimization problem is unique~\cite{Lin_PointEstimation_2018}.

However, in a real experiment, it would be hard to justify that the trials are \iid,~since this entails running every trial under the exact same conditions, which would be impractical with imperfect devices. The key observation of the PBR protocol is that even for non-\iid~trials, the following Bell-like inequality remains valid~\cite{ZGK11,LZ19} for all $\vec{P}\in\NS$:
\begin{subequations}\label{Eq:Bell-like}
\begin{equation}\label{Eq:BellExp}
    \sum_{a,b,x,y} R_{abxy}P(x,y)P(a,b|x,y) \stackrel{\NS}{\le} 1
\end{equation}
where the coefficients $R_{abxy} \ge 0$ for all $a,b,x,y$ are the so-called prediction-based ratios (PBRs), defined as,\footnote{Due to numerical imprecisions, the solver may only find a correlation close to the true minimizer $\vecPstar$. Then, the Bell-like inequality of~\cref{Eq:Bell-like} only holds approximately, with the maximum of the left-hand-side of \cref{Eq:BellExp}  over all $\vecP\in\NS$ being $1+\epsilon$, for some tiny $\epsilon>0$. In this case, we ought to renormalize (i.e., divide) the PBRs obtained from \cref{eq:PBR_bell_ineq} by $1+\epsilon$ to ensure that the $p$-value bound obtained thereafter is valid.}
\begin{equation}
\label{eq:PBR_bell_ineq}
    R_{abxy}:= \frac{f(a,b|x,y)}{P^\mathrm{NS}_\star(a,b|x,y)}.
\end{equation}
\end{subequations}
Note that \cref{Eq:Bell-like} is an optimized Bell-like inequality for witnessing the violation of the $\NS$ hypothesis by data that follows the distribution governed by $\vecf$ (see~\cite{ZGK11} for a discussion based on the hypothesis of LHV theories associated with $\L$). Hence, if the subsequent trials follow a distribution significantly different from the $\vecf$ used in defining \cref{eq:PBR_bell_ineq}, even if the data violates the $\NS$ hypothesis, it may not be reflected by the corresponding $p$-value bound determined from the above PBRs.

For the purposes of hypothesis testing via the PBR protocol, we only use part of the data to establish the Bell-like inequality of \cref{Eq:Bell-like}, while the remaining part is used to compute a test statistic from its coefficients.
Suppose we take the first $N_\mathrm{est} < N$ trials of the data to obtain $R_{abxy}$ via \cref{Eq:f,Eq:DKL,eq:PBR_bell_ineq}, i.e., the right-hand side of \cref{Eq:SumNxy} is now $N_\mathrm{est}$.
Then, we have the remaining $N_\mathrm{test}:= N - N_\mathrm{est}$ sets of data for computing a $p$-value upper bound.
Let $(x_i,y_i)$ denote the settings and $(a_i,b_i)$ the outcomes observed in the $i$-th trial.
The PBR for this round would be $r_i := R_{a_ib_ix_iy_i}$, which corresponds to the value of $R_{abxy}$ for the combination of inputs and outputs seen in the trial.
In the PBR protocol, we consider a test statistic given by the product of all $r_i$'s from the $N_\mathrm{test}$ remaining trials:
\begin{equation}
\label{eq:teststatistic}
    t = \prod_{i=N_\mathrm{est}+1}^{N} r_i
     = \prod_{a,b,x,y} R_{abxy}^{N_{a,b,x,y}},
\end{equation}
where $N_{a,b,x,y}$ is now the number of times the combination $(a,b,x,y)$ occurs in the $N_\mathrm{test}$ hypothesis-testing trials.

Let $T_m$ denote the random variable obtained from the product of the PBRs of $m$ trials.
It can be shown~\cite{ZGK11} that if each $r_i$ satisfies \cref{eq:PBR_bell_ineq}, then we have that $\mathbb{E}(T_{i+1}|H_{\le i}) \le \mathbb{E}(T_{i})$, where $\mathbb{E}$ denotes the expectation value and $H_{\le i}$ denotes all past information obtained until the $i$-th trial.
This means the probability that $T_m$ exceeds a particular value $t$ can be upper bounded using Markov's inequality, and the upper bound itself is our $p$-value upper bound $p_U$:
\begin{equation}
    \Pr[T_{N_\text{test}} \ge t ] \le \min\left( t^{-1}, 1 \right) =: p_U.
\end{equation}
A small $p_U$, and hence a small $p$-value, would represent a large value of $t$, which would only occur if we had sufficiently many $r_i > 1$. Note that this argument relies only on the supermartingale property of $T_m$, thus anything we conclude from the hypothesis testing is valid even with non-\iid~trials. In contrast, if $t^{-1}>1$, we have the {\em trivial} $p$-value bound $p_U=1$, which does not provide any useful information about the validity of the null hypothesis.

Before presenting our results and analysis, let us briefly comment on one final subtlety regarding the detection of NS violation via a Bell-like inequality violation. Clearly, the NS constraints of \cref{Eq:NS} consist of a collection of equality constraints.
To see their connection with an inequality like~\cref{Eq:Bell-like}, it suffices to remember that any equality constraint ($=$) is {\em equivalent} to the conjunction of two inequality constraints ($\ge$ and $\le$). In other words, violating any of the NS conditions must also imply a violation of at least one inequality constraint analogous to those shown in~\cref{Eq:NS}.

\section{CHSH Bell tests in IBM quantum computers}\label{App:IBM}

As mentioned at the beginning of \cref{Sec:Sets}, in this work, we focus on the case where $|X|=|Y|=|A|=|B|=2$. In this case, it is known~\cite{Brunner2014:RMP} that $\L$ can be equivalently specified as the intersection of positivity facets and eight different versions of the Clauser-Horne-Shimony-Holt (CHSH)~\cite{CHSH1969} Bell inequality.  In what follows, we explain how the PBR protocol can be applied to the data collected in this simplest Bell scenario in conjunction with various hypotheses that allow us to identify measurement cross-talks.
Note, however, that the analysis can be easily adapted to other certification tasks when the NS conditions of \cref{Eq:NS} hold and more complicated Bell scenarios, as illustrated in~\cite{CCC+24}.

For the CHSH Bell test on an IBMQ device, we consider the setting where Alice and Bob share a two-qubit state and they perform a local measurement in two possible bases on each qubit. In the actual implementation, this means we first choose the pair of qubits representing Alice and Bob.
Then, each round of the Bell test goes as follows: First, we apply the quantum gates needed to prepare the initial shared state. Next, according to the pair of inputs $(x,y)$ with $x,y\in\{0,1\}$, we perform one of the four possible quantum circuits that implement the local measurements to obtain the pair of outcomes $(a,b)$. We record the data $(a,b,x,y)$ in each round to facilitate subsequent analysis.

In a typical Bell test, one enforces the NS conditions in one way or another and seeks to demonstrate Bell nonlocality. Here, we test whether the observations are consistent with a quantum model {\em assuming} local, independent measurements, \cref{eq:quantumP}, or more generally, the NS constraints of~\cref{Eq:NS}.
To this end, we focus on the commonly encountered Bell test that aims to produce a Bell-nonlocal correlation maximally violating the CHSH Bell inequality.
We also fix Alice's and Bob's two measurements to be the ones in the computational (Pauli-$Z$) and Hadamard (Pauli-$X$) bases for $x,y=0,1,$ respectively.
The Bell test can then be conveniently described using the quantum circuits we implement on the IBMQ devices.

\begin{figure}[h!tbp]
    \centering    \includegraphics[width=0.765\columnwidth]{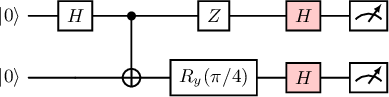}
\caption{Quantum circuits generating (ideally) the maximal CHSH-Bell-inequality-violating correlation. $H$ is the Hadamard gate, $R_y(\theta) = \cos\frac{\theta}{2}\mathbb{1} - i\sin\frac{\theta}{2} Y$, where $\mathbb{1}$ and $Y$ are, respectively, the identity and the Pauli-$Y$ operator, and the meter symbol represents the computational basis measurement. The pink shading indicates that the Hadamard gate is implemented before the measurement only when the input for the top (bottom) qubit is $x= 1$ ($y = 1$).}
\label{Eq:CNL}
\end{figure}

Specifically, the circuits $\CirNL$ to generate a Bell-nonlocal correlation are given in~\cref{Eq:CNL}. Ideally, this circuit prepares the maximally entangled state
\begin{equation}
    \ket{\psi} = \frac{1}{\sqrt{2}} \left[\cos\frac{\pi}{8}(\ket{00}-\ket{11})+\sin\frac{\pi}{8}(\ket{01}+\ket{10}) \right],
\end{equation}
and measures Pauli-$Z$ and Pauli-$X$ on both qubits, thereby giving the maximal-CHSH-violating nonlocal correlation.

Now that we have specified the Bell test, it remains to choose the two specific qubits in the IBMQ device to represent Alice and Bob.
To demonstrate the viability of the PBR protocol, we use the information about the average CNOT gate errors reported around the last week of April 2023 in several IBMQ devices to select those pairs of qubits with relatively high errors, see~\cref{App:Details} for details.  The pairs of qubits chosen are indicated in~\cref{tab:ibmq_qubits} using the device name and qubit numbers.

\begin{table}[ht!]
    \centering
    \begin{tabular}{|c|c|}
    \hline
        IBMQ device & Qubit Pairs  \\
        \hline
         Washington & (12,17)\, (38,39)\, (79,91)\, (91,98) \\
         Geneva & (7,10)\, (14,16)\, (21,23) \\
         Cairo  & (0,1)\, (7,10)\, (13,14)\, (23,24) \\
         Hanoi  & (5,8)\, (6,7)\, (11,14)\, (19,20) \\
         Mumbai & (5,8)\, (16,19)\, (23,24) \\
         \hline
    \end{tabular}
    \caption{List of qubit pairs in each IBMQ device where we perform the two types of CHSH Bell tests. For example, Washington(12,17) means the qubit pair $(12,17)$ of the IBMQ device Washington. For the topology of the qubit connections in these devices and their calibration data, see~\cref{App:Details}.}
    \label{tab:ibmq_qubits}
\end{table}

A few remarks on the data acquisition process are now in order.
In an IBMQ device, a task consists of specifying the quantum circuit to be implemented and the number of shots, i.e., how many times we repeat the experiment.
However, for a proper Bell test, the inputs $(x,y)$ must be generated randomly and uncorrelated with the state of the qubits to be tested. To this end, one may first generate a (pseudo)random sequence of input pairs $(x,y)$ and submit a task defined by the sequence of circuits corresponding to these pairs while setting the number of shots to {\em unity} for each circuit.

\begin{figure}[t!]
    \centering    \includegraphics[width=1\columnwidth]{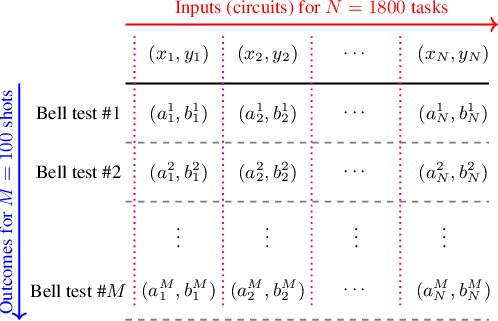}
    \caption{Schematic showing how the data for hypothesis testing is collected in each IBMQ device. For each device, we submit $N=1800$ tasks with $M=100$ shots each. The $k$-th task is specified by the two input bits $(x_k,y_k)$, corresponding to one of the four circuits shown in \cref{Eq:CNL}. All results $(a_k^i,b_k^i)$ for the $i$-th shot across the different tasks are then consolidated as results from the $i$-th Bell test on each device. }
    \label{fig:DataAcquisition}
\end{figure}

Even then, the issue remains that various shots may become correlated since we must specify the entire input bit strings $(x_1,y_1),(x_2,y_2),\cdots,(x_N,y_N)$ when submitting the task. In other words, from the perspective of a loophole-free Bell test~\cite{Hensen2015,Shalm2015,Giustina2015,rosenfeld_event-ready_2017}, this potentially allows the leakage of inputs across parties. For cross-talk detection, we shall {\em assume} that this potential leakage does not alter the behavior of the individual qubits. Even though this {\em assumption} renders our protocol {\em non-fully-DI}, and hence our choice of the term {\em almost DI},\footnote{The term almost device-independent was also used very differently in \cite{SOS+24} to refer to a situation where only one of the parties in a multipartite scenario is trusted.} its violation would again imply some kind of cross-talks that should be addressed to improve local addressability. Moreover, to collect statistically significant sets of data more efficiently, instead of measuring one shot for each circuit, we carry out multiple shots for each circuit but assign each shot to a different Bell test.
This means if we want to run $M$ Bell tests where each Bell test consists of $N$ rounds (experimental trials), we submit $N$ tasks to an IBMQ device where each task consists of $M$ shots. Then, the data produced by the $i$-th shot of every task is treated as the data for the $i$-th Bell test. See \cref{fig:DataAcquisition} for a schematic explanation of the data acquisition process.

Finally, to make the comparisons across different IBMQ devices relatively fair, we standardize each CHSH Bell test of \cref{Eq:CNL} to have $N = 1800$ trials, and we perform $M=100$ tests for each pair of qubits chosen. Moreover, for the PBR analysis, we use the data from the first $N_\mathrm{est} = 600$ trials to obtain the empirical frequencies $\vec{f}$, and the remaining $N-N_\mathrm{est} =1200$ trials for computing the $p$-value upper bounds. Importantly, one can equally well make other choices of $N_\mathrm{est}$. The general principle here is that we need a sufficient amount of data to get a reasonably good estimate of the general behavior (via $\vec{f}$), and hence a good PBR via~\cref{eq:PBR_bell_ineq},  but we also need a sufficient amount of data from {\em different} set of trials for performing the actual hypothesis testing (via the test statistic). In our analysis, we adopt a significance level of $\alpha = 0.05$, which means we reject the null hypothesis if the $p$-value bound is less than $\alpha$.

\section{Results}
\label{Sec:Results}

After collecting the data from the Bell tests described in the previous section, we perform various PBR analyses by testing the data against different null hypotheses.

\subsection{PBR protocol for revealing the violation of Born's rule for local measurements}
\label{Sec:BornViolation}

Since we are interested in the local addressability of these devices, we start by employing a PBR analysis to check for signatures that the measurement statistics violate Born's rule for local measurements, cf. \cref{eq:quantumP}. Due to statistical fluctuations, empirical frequencies $\vec{f}$ of \cref{Eq:f} typically do not satisfy the NS constraints of \cref{Eq:NS}. While this does not compromise the PBR protocol in computing valid $p$-value bounds, previous studies~\cite{ZGK11,ZGK13,CCC+24} have suggested that the quality of $p$-value bounds may be improved by first transforming $\vec{f}$ into an initial estimate $\vec{G} \in \NS$. For example, we can set~\cite{Lin_PointEstimation_2018} $\vec{G}$ as the minimizer of the KL-divergence from $\vec{f}$ to $\NS$:
\begin{equation}
	\vec{G}:=\argmin_{\vec{P} \in \NS} \DKL(\vec{f}||\vec{P}).
\end{equation}
The initial estimate $\vec{G}$ then plays the role of the frequencies $\vec{f}$ in the subsequent PBR analysis in \cref{Eq:DKL,Eq:Bell-like}.

More precisely, in determining the PBR for the hypothesis of \cref{eq:quantumP} via \cref{Eq:DKL},  we use $\Q^3$, level-3 of the Moroder hierarchy~\cite{Moroder13} as a proxy for the local quantum constraints of \cref{eq:quantumP}, see the last paragraph of \cref{Sec:Sets}. Hence, our null hypothesis, in fact, corresponds to the set $\Q^3$, which is a strict outer approximation of $\Q$. Still, a small $p$-value bound for $\Q^3$ signifies the violation of~\cref{eq:quantumP}, in the sense that a rejection of $\vecP\in\Q^3$ must entail a rejection of $\vecP\in\Q$ since $\vecP\not\in\Q^3\implies \vecP\not\in\Q$, see~\cref{Fig:Inclusions}.

After the estimation stage, we obtain the PBRs
\begin{equation}
\label{eq:PBRwithInitialEst}
    R_{abxy} = \frac{G(a,b|x,y)}{P_\star^{\Q^3}(a,b|x,y)},
\end{equation}
where $\vec{P}_\star^{\Q^3}$ represents the minimizer of the KL-divergence from $\vec{G}$ to $\Q^3$, i.e.,
\begin{equation}\label{Eq:Q3Minimizer}
    \vec{P}_\star^{\Q^3}:=\argmin_{\vec{Q}\in\Q^3} \DKL(\vec{G}||\vec{Q}).
\end{equation}
Next, we proceed to the hypothesis testing stage using the PBRs $R_{abxy}$ from~\cref{eq:PBRwithInitialEst} in the usual way. That is, we compute the test statistic in \cref{eq:teststatistic} from the number of occurrences  $N_{a,b,x,y}$ of the input-output combination $(a,b,x,y)$ in the hypothesis testing trials.

\begin{table}[t!]
    \centering
    \begin{tabular}{|c|c|c|c|c|c|c|}
    \hline
    Device & Qubits [Circuit] & $\Q^3$ & $\NS$ & $\NSnotAtoB$ & $\NSnotBtoA$ & $\L$    \\
    \hline
    \multirow{4}{*}{Washington}
    &  12,17 [$\CirNL$] & 1 & 1 & 0 & 1 & 0 \\
    &  38,39 [$\CirNL$] & 2 & 2 & 0 & 1 & 2 \\
    &  79,91 [$\CirNL$] & 1 & 1 & 0 & 0 & 1 \\
    &  91,98 [$\CirNL$] & 2 & 2 & 1 & 1 & 1 \\
    \hline
    \multirow{2}{*}{Geneva} &
    14, 16 [$\CirNL$] & 0 & 0 & 0 & 1 & 0 \\
    &  21, 23 [$\CirNL$] & 19 & 19 & 5 & 30 & 17 \\
    \hline
     Cairo & 13,14 [$\CirNL$] & 1 & 1 & 1 & 0 & 100 \\
    \hline
    \multirow{3}{*}{Hanoi}
    &  5,8 [$\CirNL$]   & 2 & 2 & 0 & 1 & 47 \\
    &  11,14 [$\CirNL$]  & 0 & 0 & 0 & 2 & 100 \\
    &  19,20 [$\CirNL$] & 1 & 2 & 0 & 0 & 99 \\
    \hline
    Mumbai &
       23,24 [$\CirNL$] & 1 & 1 & 0 & 1 & 64 \\
    \hline
    \end{tabular}
    \caption{Summary of nontrivial hypothesis-testing results based on the PBR protocol applied to the data collected in Bell tests performed on various IBMQ devices via the circuits $\CirNL$ of \cref{Eq:CNL} during 2023-04 to 2023-05 (see \cref{tab:DataDates} in \cref{App:Details} for details).  For each qubit pair, we implement 1800 tasks with 100 shots each, which means we conduct $M=100$ separate Bell tests with $N=1800$ trials each.  For each Bell test, we run the PBR protocol for various hypotheses $\H\in\{\Q^{3},\NS,\NSnotAtoB,\NSnotBtoA,\L\}$ with $N_\mathrm{est} = 600$ at a significance level of $\alpha = 0.05$. The integers from the third to the rightmost column show the number of Bell tests where we observe a signature, with a confidence of at least $95\%$, for the violation of various hypotheses: a relaxation of Born's rule for local measurements ``$\Q^3$'', (two-way) no-signaling ``$\NS$'', no-signaling from $A$ to $B$ ``$\NSnotAtoB$", no-signaling from $B$ to $A$ ``$\NSnotBtoA$'', and LHV ``$\L$''. Qubit $A$ ($B$) corresponds to the first (second) integer entry in the second column. Only the combinations of device and qubit-pair where at least one of the entries from the third to the sixth column is nonzero is listed. For the corresponding results with the significance level tightened to $\alpha=0.01$, see \cref{tab:smallPval001} in \cref{App:MiscResults}.}
    \label{tab:smallPval}
\end{table}

Finally, for the IBMQ devices listed in \cref{tab:ibmq_qubits} and each of the $M=100$ Bell tests performed, we compare the $p$-value upper bound obtained against the significance level $\alpha = 0.05$ to decide if the null hypothesis corresponding to (a relaxation of) \cref{eq:quantumP} should be rejected. In \cref{tab:smallPval}, we list the IBMQ devices (alongside the qubit pairs) where at least one instance of rejection is recommended by the PBR protocol for this significance level. In most cases, we only observe one or two such instances. However, for our implementation of the circuits of \cref{Eq:CNL} using qubit pair Geneva(21,23), cf.~\cref{Fig:Geneva}, we find, with a confidence of at least $95\%$, incompatibility with \cref{eq:quantumP} for $19$ out of $100$ instances of the conducted Bell tests. A histogram showing the distribution of these $p$-value bounds can be found in~\cref{fig:Geneva21_23_Q3}. These results reveal a strong signature for the inappropriateness of using \cref{eq:quantumP} to model the measurement statistics on these two qubits of this particular IBMQ device.

\subsection{PBR protocol for revealing signaling effects}

Since the measurement results analyzed above are those generated from the circuits of \cref{Eq:CNL}. Their incompatibility with \cref{eq:quantumP} even when non-\iid~behavior is allowed, i.e., the state $\rho_{AB}$ in \cref{eq:quantumP} is allowed to vary from one trial to another), is already a strong indication that cross-talks are present in some of these IBMQ devices. A more direct evidence of this undesired aspect follows if we can show that the measurement statistics exhibit signaling effects, i.e., violate {\em one or more} of the NS conditions given in~\cref{Eq:NS}. To this end, instead of following the analysis presented in~\cref{Sec:BornViolation}, we proceed according to the illustration given in \cref{Sec:PBR} to obtain $p$-value upper bounds according to the NS hypothesis. The corresponding results are listed under column $\NS$ of \cref{tab:smallPval}.

\begin{figure}[t!]
    \centering    \includegraphics[width=\columnwidth]{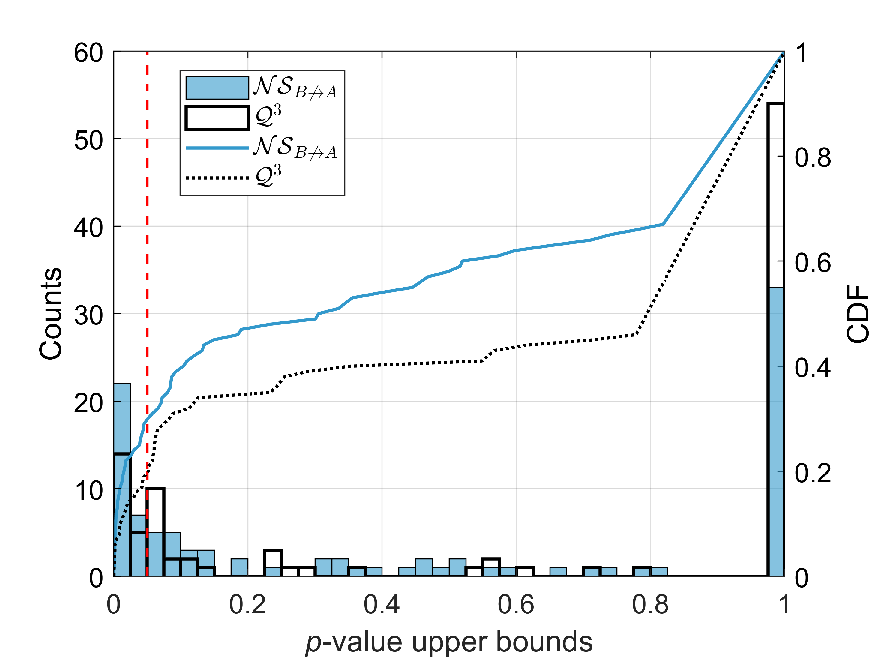}
    \caption{Histogram (H) and cumulative distribution function (CDF) of $p$-values upper bounds obtained from the PBR protocol for qubits 21 and 23 of the IBMQ device Geneva (see~\cref{Fig:Geneva}) for hypotheses $\H = \Q^3$ (H: unfilled, thicker edged boxes; CDF: dotted black line) and $\H=\NSnotBtoA$ (H: filled blue boxes; CDF: solid blue line). With the inclusion relations of \cref{Eq:Inclusions}, see also \cref{Fig:Inclusions}, rejecting the latter hypothesis for a particular test must also entail a rejection of the $\NS$, and hence the $\Q^3$ hypothesis for the same set of data. Each bin spreads over a $p$-value bound of 0.025. The dashed red vertical line indicates the significance level of $\alpha = 0.05$.
    }
    \label{fig:Geneva21_23_Q3}
\end{figure}

When we compare these results with the ones obtained above for the $\Q^3$ hypothesis (see also~\cref{tab:smallPval:details}), we see that they are almost identical, except for the Bell test corresponding to the $21$st shots on Hanoi(19,20) [\cref{Fig:Hanoi}], where we find evidence for the violation of $\NS$ but not $\Q^3$. In this case, we find a $p$-value bound of $0.046$ for the former but only the trivial bound of $1$ for the latter. Since $\Q^3\subset\NS$, i.e., $\NS$ is a less-constraining set of correlations than $\Q^3$, the above observation may appear counterintuitive at first glance, as one may expect to see fewer, rather than more, instances of violation of the $\NS$ hypothesis.
To understand the origin of this discrepancy, we remind that our protocol involves an estimation of the PBR (i.e., a Bell-like inequality) optimized for the relative frequencies $\vecf_\text{est}$ deduced from the data collected during the first $N_{\rm est}$ trials in each Bell test. However, for finite and non-\iid~trials, there is no guarantee that such an estimate is again optimal for the data collected during the remaining trials, cf.~\cite{ZGK11}. Indeed, for this specific Bell test, if we had used the PBR obtained for $\NS$---also valid for the $\Q^3$ hypothesis---for our test against $\Q^3$, we would have also concluded a rejection of $\Q^3$.

In other words, if the underlying process is always described by {\em a fixed} $\vecP\not\in\NS$, we must also have $\vecP\not\in\Q^3$. However, for non-\iid~trials, as remarked in the paragraph below~\cref{eq:PBR_bell_ineq}, if the empirical frequencies $\vecf_\text{est}$ do not reflect well the behavior of subsequent trials, the PBR derived therefrom for $\Q^3$ may fail to manifest the incompatibility between the hypothesis-testing trials data and $\Q^3$. In contrast, even if $\vecf_\text{test}$ differ considerably from $\vecf_\text{est}$, so long as the main signaling direction is preserved, it is conceivable that the PBR derived for $\NS$ remains effective for the testing trials.

\begin{table}[t!]
    \centering
    \begin{tabular}{|c|c|c|c|c|c|c|}
    \hline
    Device & Qubits [Circuit] & $\Q^3$/ $\NS$  & $\NSnotAtoB$ & $\NSnotBtoA$     \\
    \hline
    \multirow{4}{*}{Washington}
    &  12,17 [$\CirNL$] & 56  & - & 49  \\
    &  38,39 [$\CirNL$] & 88, 100  & - & 88  \\
    &  79,91 [$\CirNL$] & 41  & - & -  \\
    &  91,98 [$\CirNL$] & 15, 50  & 15 & 15  \\
    \hline
    \multirow{6}{*}{Geneva} &
    14, 16 [$\CirNL$] & -  & - & 99  \\
    &   \multirow{5}{*}{21, 23 [$\CirNL$]}  & $\mathfrak{C}$, 9,  &   & $\mathfrak{C}$, 6, 33, 39  \\
    & & 17, 26,   & 9, 29,   & 41-44, 47\\
    & & 32, 51,   &38, 66,  & 56-59, 68\\
    & & 66, 73,   & 73 & 72, 86, 89\\
    & & 99 & &  90, 92, 94\\
    \hline
     Cairo & 13,14 [$\CirNL$] & 8  & 8 & -  \\
    \hline
    \multirow{3}{*}{Hanoi}
    &  5,8 [$\CirNL$]   & 23, 52  & - & 23  \\
    &  11,14 [$\CirNL$]  & - & -  & 65, 75  \\
    &  19,20 [$\CirNL$] & 75 (21) & - & -  \\
    \hline
    Mumbai &
       23,24 [$\CirNL$] & 70  & - & 70  \\
    \hline
    \end{tabular}
    \caption{Further details about the instances of Bell tests giving the results reported in \cref{tab:smallPval}. Under the third column to the rightmost, we list the Bell test number implemented on the respective device and qubit pair that shows a violation of the corresponding null hypothesis. To simplify the presentation, we have put the almost identical results for $\Q^3$ and $\NS$ under the same column, with the additional instance for $\NS$ in a bracket. Moreover, we denote the common instances for $\Q^3$ ($\NS$) and $\NSnotBtoA$ in the case of Geneva(21,23) by $\mathfrak{C}= 3, 16, 18, 19, 31, 38, 55, 62, 65, 81, 97$.
}
\label{tab:smallPval:details}
\end{table}

Apart from this one exceptional instance with Hanoi(19,20), the compatibility of every other Bell test's data with the two hypotheses (i.e., whether the $p$-value bound is less than $\alpha$) is the same. In fact, even though the two hypotheses are not the same, the difference in their $p$-value bounds is typically not large enough to alter their distribution in a significant manner. For example, for the $p$-value upper bounds shown in \cref{fig:Geneva21_23_Q3}, the corresponding $p$-value bounds for the $\NS$ hypothesis differ from the former by at most $0.0017$ and are thus not visibly different from the histogram of \cref{fig:Geneva21_23_Q3} for $\Q^3$ (unfilled, thicker edge).

While small $p$-values indicate strong evidence against the $\NS$ hypothesis, they do not tell us anything about how the NS constraints of~\cref{Eq:NS} are violated.
One possibility is that including a Hadamard or not before the top (bottom) qubit measurement in \cref{Eq:CNL} indeed results in different measurement statistics on the other qubit, which, of course, goes against the assumption of \cref{Eq:NS}, and hence \cref{eq:quantumP}.
To this end, it will also be useful to check if the cross-talk has a specific directionality by running a PBR protocol
assuming the hypothesis $\mathcal{H} = \NSnotAtoB$ ($\mathcal{H} = \NSnotBtoA$) of one-way no-signaling from Alice to Bob (Bob to Alice).
Our results for these tests can be found in their respective columns in~\cref{tab:smallPval}.

From \cref{tab:smallPval,tab:smallPval:details}, we observe several instances---namely, Hanoi(11,14), Geneva(14,16), and Geneva(21,23)---where more violation of the {\em less} constraining OWNS hypothesis (either $\NSnotAtoB$ or $\NSnotBtoA$) is observed, but via the PBR protocol described in \cref{Sec:BornViolation}, fewer or {\em no} violation of the {\em more} constraining $\NS$ hypothesis is picked up. This anomaly can again be understood from the non-\iid~nature of the experimental trials, where the main direction of signaling (estimated from $\vecf_\text{est}$ and $\vecf_\text{test}$) changes from the first 600 trials to the remaining 1200 trials.

In fact, we can ``utilize'' this discrepancy to our advantage in our hypothesis-testing tasks. By recalling from \cref{Eq:Inclusions} and \cref{Fig:Inclusions} the strict inclusions of the various sets of correlations, we note that $\Q$ is the most constraining hypothesis among all those discussed above, while the OWNS hypothesis is the weakest. In other words, if we reject the plausibility of any of the hypotheses from $\{\NSnotAtoB,\NSnotBtoA\}$ in explaining the data observed for a particular Bell test, we must also reject the plausibility of $\NS$ (and hence $\Q$) in explaining the same set of data. Using this observation, we conclude from~\cref{tab:smallPval:details} that of the 100 tests performed on Geneva(21,23) 39 are deemed incompatible with the no-signaling hypothesis $\NS$ (or $\Q$). See~\cref{tab:Combined} for a complete summary of such results on all the IBMQ devices we have tested.

\section{Discussion}~\label{Sec:Discussion}

In recent years, due to the widespread availability of quantum computers through the cloud, we have seen a surging interest in running various quantum tasks on these devices. Naturally, given the proximity of the qubits arranged in some of these platforms---such as those offered by IBM Quantum (IBMQ)---one may wonder about the extent to which they exhibit cross-talks and whether such effects can detected with minimal assumptions, like other device-independent (DI) certification tasks. To this end, it is worth noting that the no-signaling (NS) conditions of \cref{Eq:NS} are usually separately enforced and taken as a premise for DI protocols.

In this work, we show under a mild assumption that measurement cross-talks or incompatibility with Born's rule for local measurements can again be certified in an essentially DI manner via the PBR protocol (initially developed in~\cite{ZGK11,ZGK13} for testing LHV theories but later generalized in~\cite{LZ19}). More precisely, we use the protocol to obtain $p$-value upper bounds on the plausibility of the NS assumption or the natural assumption that Born's rule for local measurements holds. Note that an analysis of the first kind has previously been applied as a consistency check in the loophole-free Bell test performed with superconducting circuits~\cite{Storz:2023aa}, where no evidence for signaling is found.

\begin{table}[t!]
    \centering
    \begin{tabular}{|c|c|c|c|}
    \hline
    Device & Qubits[Circuit] & $\alpha=0.05$ & $\alpha=0.01$   \\
     \hline
    \multirow{4}{*}{Washington} &
      12,17 [$\CirNL$] & 2 & 1 \\
    &  38,39 [$\CirNL$] & 2 & 1 \\
    &  79,91 [$\CirNL$] & 1 & 0 \\
    &  91,98 [$\CirNL$] & 2 & 1\\
    \hline
    \multirow{2}{*}{Geneva} &
    14, 16 [$\CirNL$] & 1 & 1 \\
    &  21, 23 [$\CirNL$] & 39 & 22\\
    \hline
     Cairo & 13,14 [$\CirNL$] & 1 & 1 \\
    \hline
    \multirow{3}{*}{Hanoi} &
      5,8 [$\CirNL$]   & 2 & 1 \\
   & 11,14 [$\CirNL$]   & 2 & 0 \\
   & 19,20 [$\CirNL$]   & 2 & 0 \\
    \hline
     Mumbai & 23,24 [$\CirNL$] & 1 & 0 \\
    \hline        \hline
    \multirow{4}{*}{Washington} &
      12,17 [$\CirL$] & 1 & 0 \\
    &  38,39 [$\CirL$] & 2 & 1 \\
    &  79,91 [$\CirL$] & 2 & 1 \\
    &  91,98 [$\CirL$] & 2 & 1\\
    \hline
    \multirow{2}{*}{Cairo} &
      13,14 [$\CirL$]   & 2 & 0 \\
   & 23,24 [$\CirL$]   & 3 & 1 \\
   \hline
    \multirow{4}{*}{Hanoi} &
      5,8 [$\CirL$]   & 2 & 0 \\
   & 6,7 [$\CirL$]   & 1 & 0 \\
   & 11,14 [$\CirL$]   & 2 & 1 \\
   & 19,20 [$\CirL$]   & 1 & 0 \\
    \hline
    \end{tabular}
    \caption{Summary of the number of (nonzero) instances of Bell tests found to be incompatible with the no-signaling hypothesis, either via the rejection of the $\NS$ null hypothesis, or indirectly via one of the OWNS hypotheses for a significance level of $5\%$ (third column) and $1\%$ (fourth column). We find the same results for rejecting the hypothesis of Born's rule for local measurements, \cref{eq:quantumP}. Results listed on top and bottom are, respectively, those based on the circuits $\CirNL$ of \cref{Eq:CNL} (meant for generating a Bell-nonlocal correlation) and $\CirL$ of \cref{Eq:CL} (meant for generating a Bell-local correlation, see~\cref{App:Local} for details).}
    \label{tab:Combined}
\end{table}

Similarly, from our analysis of the data obtained across five different IBMQ systems, we see, in most cases, very little evidence for a strong violation of either the $\Q^3$ or any of the $\NS$ hypotheses. Although we observe a small $p$-value upper bound $p_U$ in a few instances (see \cref{tab:Combined} for a summary), it should be reminded that even when the null hypothesis holds, there remains a small chance ($<p_U$) of observing a false positive~\cite{ZGK11}. In contrast, for measurements on qubits 21 and 23 of the IBMQ-Geneva device, we have stumbled upon $39$ instances of these tests where the PBR protocol would end up rejecting the $\NS$, and hence $\Q$ hypothesis, either directly, or indirectly via a weaker hypothesis.
This shows that, despite the relatively small number of samples (1800 trials for each test) and allowing non-\iid~trials (cf. the approach by~\cite{Rybotycki:2025ab} with \iid~assumption), the PBR protocol is capable of detecting (measurement) cross-talks in a real quantum computer.

Note further that when we check the same set of data from Geneva(21,23) against the $\L$ hypothesis of LHV theories,\footnote{To check against $\L$ using the PBR protocol, we replace, in the definition of the PBRs of \cref{eq:PBRwithInitialEst}, the denominator by $\argmin_{\vec{Q}\in\L} \DKL(\vec{G}||\vec{Q})$, i.e., the minimizer of the KL divergence from $\vec{G}$ to $\L$.} we also find several instances that result in rejecting the $\L$ hypothesis. However, given the observed signaling effects, the relevance of this violation becomes questionable. For comparison, we have also implemented several trivial ``Bell tests'' using the circuits $\CirL$ of \cref{Eq:CL} in \cref{App:Local}, which are only expected to produce Bell-local correlations. Then, for ideal devices, we anticipate many small $p$-values for Bell tests involving $\CirNL$, \cref{Eq:CNL}, and none for those involving $\CirL$. The results shown in \cref{tab:smallPval} and \cref{tab:smallPvalL} clearly do not follow this intuition. In fact, from \cref{tab:smallPvalL,tab:smallPvalL:details}, we even observe a few instances of rejection of $\L$ alongside $\Q^3$ and $\NS$ with $\CirL$, suggesting that these violations of~\cref{eq:localP} are merely an artifact of the cross-talks present in the system. Even though we have not seen overwhelming instances of rejections of the $\NS$ or any of the OWNS hypotheses for the $\CirL$ circuit, cf.~\cref{tab:Combined,tab:smallPvalL}, their presence, nonetheless, support the idea that these cross-talks show up even without implementing any nonlocal unitary gate.

\begin{figure}
    \centering    \includegraphics[width=0.95\columnwidth]{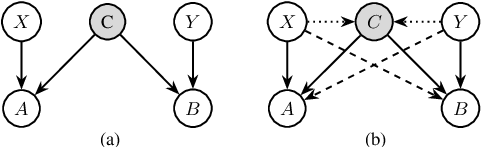}
\caption{Causal structures relevant to the current work. (a) The Bell directed acyclic graph~\cite{Wood:2015aa} (DAG) where the random variable $A$ ($B$) is allowed to depend on both $X$ ($Y$) and the latent variable $C$. When $C$ is a classical or a quantum common cause, the resulting correlation $\vecP$ satisfies, respectively, ~\cref{eq:localP,eq:quantumP}. Even if $C$ is a more powerful no-signaling common cause, $\vecP$ must still respect \cref{Eq:NS}. (b) A natural extension of the Bell DAG allowing signaling (dashed arrows) and ``measurement dependence''~\cite{PRB+14} (dotted arrows).}
\label{Fig:CausalStructure}
\end{figure}

Let us make a few final remarks about our general methodology. Our certification protocol can be seen as assuming the causal structure of~\cref{Fig:CausalStructure}(a) and applying the PBR method to show that the observed data is incompatible with this assumption.
Strictly, a refutation of the assumed causal structure does not necessarily entail signaling, and hence a measurement cross-talk. For example, the causal structure depicted in \cref{Fig:CausalStructure}(b)---which allows signaling (dashed arrows) and dependence of $C$ on $X,Y$ (dotted arrows)---
facilitates the generation of all possible $\vecP$ in this scenario, cf. Lemma 1 of~\cite{Evans:2016aa}.

In an ordinary Bell test, one invokes the freedom of choice assumption to remove these dependencies between $C$ and $X,Y$, making $P(X,Y|C)=P(X,Y)$, which is equivalent to $P(C|X,Y)=P(C)$. In our case, however, the inputs $X, Y$ are not only generated before $C$, but are even fed into the device used to prepare $C$. Thus, although we make the same independence assumption, some may find it more difficult to justify in the present context. To this end, one might find a random permutation of the bit strings (and qubits) before each submission helpful for breaking any accidental correlations between the input pattern and the underlying time-dependent noise. Alternatively, one can follow the approach of~\cite{PRB+14} and try to develop a more general type of NS conditions that hold even if we allow these undesired dependencies.

On the other hand, most of our tests admittedly involve qubit pairs exhibiting relatively high error rates. However, even among those pairs where the error rates seem low, including Washington(38, 39) (see~\cref{Fig:Washington}), Cairo(13, 14) and Cairo(23, 24)  (see~\cref{Fig:Cairo}), and Hanoi(6,7) (see~\cref{Fig:Hanoi}), our protocol has also identified instances showing signatures of cross-talk, see~\cref{tab:Combined}. For future reference,  it would be helpful to perform a comprehensive investigation involving a control set of low-error pairs and compare the results obtained against IBMQ's calibration data. In particular, this will shed light on the effectiveness of our tests---which involve {\em far fewer assumptions}---even in those cases that may not be flagged via the conventional approach.

Note also that our results (see~\cref{tab:smallPval,tab:smallPval:details,tab:smallPvalL,tab:smallPvalL:details}) clearly suggest that, once we have done the much faster computation checking against the $\NS$ hypothesis, the computation using any approximation of $\Q$ may well be redundant for detecting cross-talk. A natural question that follows is whether this observation holds in general. Another obvious question that follows is: for the same number of trials, whether one can obtain---for the sake of detecting cross talks---a tighter $p$-value bound for refuting the hypothesis of Born's rule for local measurements, \cref{eq:quantumP}, or even no-signaling of \cref{Eq:NS}. To this end, we remind the readers that the PBR protocol is only known to be optimal in the asymptotic setting (and when the trials are~\iid). For example, is there a way to adopt the analysis from~\cite{Elkouss:2016aa} to the present setting by considering the conjunction of all inequalities equivalent to~\cref{Eq:NS}?  Evidently, it is also relevant to understand if adapting the present analysis can give a useful quantification of cross-talks. Finally, given the current findings, one can ask if other, more general (almost) DI certification or calibration tasks can be developed to detect other non-desired behavior of quantum devices.

\begin{acknowledgments}

We thank Yi-Te Huang for his help in implementing some of the earlier computations at IBMQ and are very grateful to him, Marina Maciel Ansanelli, and Yanbao Zhang for many helpful discussions. We are also very grateful to anonymous reviewers for providing very helpful comments and suggestions on an earlier version of this paper. This work is supported by the National Science and Technology Council, Taiwan (Grants No.~109-2112-M-006-010-MY3,112-2628-M006-007-MY4,113-2918-I-006-001), the Foxconn Research Institute, Taipei, Taiwan, and in part by the Perimeter Institute for Theoretical Physics. Research at Perimeter Institute is supported by the Government of Canada through the Department of Innovation, Science, and Economic Development, and by the Province of Ontario through the Ministry of Colleges and Universities.

\end{acknowledgments}

 \appendix

\section{Miscellaneous details }\label{App:Details}

\begin{figure}[h]
    \centering    \includegraphics[width=0.8\columnwidth]{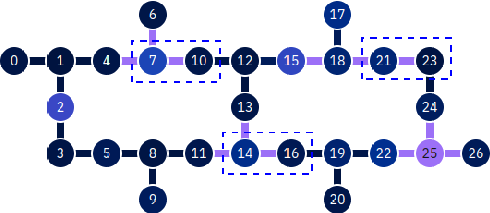}
    \caption{Topology of the 27-qubit exploratory IBM Geneva device and its calibration data on 2023-04-26: the readout (CNOT) assignment error ranges from $7.300\times10^{-3}$ to $3.683\times10^{-1}$ ($3.872\times10^{-3}$ to $1.000$) with a median of $2.930\times10^{-2}$ ($5.457\times10^{-2}$). Here and below, the highest (lowest) error is associated with the brightest (darkest) color; qubit pairs analyzed in this work are enclosed in dashed blue rectangles.
    }
    \label{Fig:Geneva}
\end{figure}

\begin{figure}[h]
    \centering    \includegraphics[width=0.8\columnwidth]{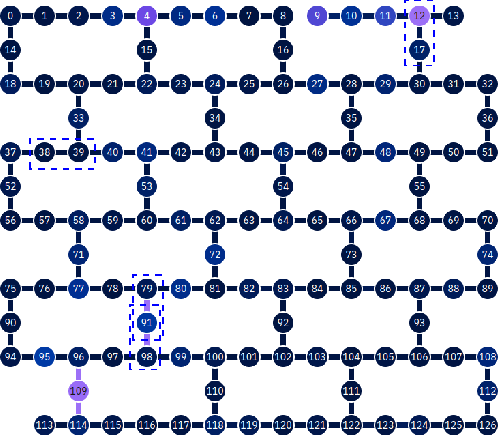}
    \caption{Topology of the 127-qubit IBM Washington device and its calibration data on 2023-04-24:  the readout assignment (CNOT) error ranges from $1.900\times10^{-3}$ to $4.854\times 10^{-1}$ ($5.999\times10^{-3}$ to $1.000$) with a median of $1.290\times10^{-2}$ ($1.234\times10^{-2}$).
    }
    \label{Fig:Washington}
\end{figure}

\begin{figure}[h]
    \centering    \includegraphics[width=0.8\columnwidth]{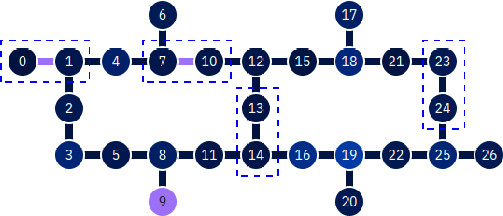}
    \caption{Topology of the 27-qubit IBM Cairo device and its calibration data on 2023-04-24: the readout assignment (CNOT) error ranges from $6.000\times10^{-3}$ to $1.221\times 10^{-1}$ ($4.436\times10^{-3}$ to $1.000$) with a median of $1.190\times10^{-2}$ ($9.815\times10^{-3}$).
}
    \label{Fig:Cairo}
\end{figure}

\begin{figure}[h]
    \centering    \includegraphics[width=0.8\columnwidth]{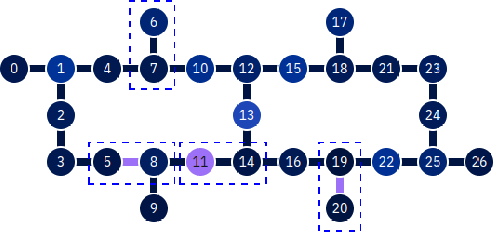}
    \caption{Topology of the 27-qubit IBM Hanoi device and its calibration data on 2023-04-24: the readout assignment (CNOT) error ranges from $5.800\times10^{-3}$ to $8.690\times 10^{-2}$ ($2.982\times10^{-3}$ to $1.000$) with a median of $1.150\times10^{-2}$ ($7.465\times10^{-3}$).
    }
    \label{Fig:Hanoi}
\end{figure}

\begin{figure}[h]
    \centering    \includegraphics[width=0.8\columnwidth]{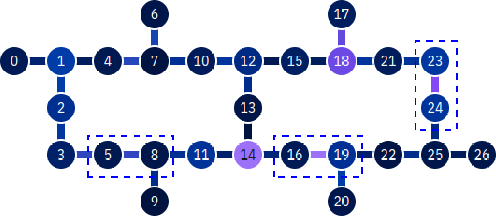}
    \caption{Topology of the 27-qubit IBM Mumbai device and its calibration data on 2023-05-08:  the readout assignment (CNOT) error ranges from $1.150\times10^{-2}$ to $1.115\times 10^{-1}$ ($4.201\times10^{-3}$ to $1.888\times10^{-2}$) with a median of $1.860\times10^{-2}$ ($7.804\times10^{-3}$).
    }
    \label{Fig:Mumbai}
\end{figure}

Here, we provide further details about the data acquisition period, \cref{tab:DataDates}, and the IBMQ devices investigated in this work. These include IBM Cairo (\cref{Fig:Cairo}), the exploratory---now retired---IBM Geneva (\cref{Fig:Geneva}), IBM Hanoi (\cref{Fig:Hanoi}), IBM Mumbai (\cref{Fig:Mumbai}), and IBM Washington (\cref{Fig:Washington}).

\subsection{Data acquisition period}

\begin{table}[h]
    \centering
    \begin{tabular}{|c|c|c|}
     \hline
         Device & Data collection period \\
          \hline
         Washington & 2023-04-24 to 2023-04-26 \\
         Geneva & 2023-04-26 to 2023-04-27  \\
         Cairo & 2023-04-24 to 2023-04-30 \\
         Hanoi & 2023-04-24 to 2023-05-17 \\
         Mumbai & 2023-05-03 to 2023-05-06\\
          \hline
    \end{tabular}
    \caption{Period for which we collected the data at IBMQ devices.}
    \label{tab:DataDates}
\end{table}

\subsection{Topology of qubit connections in each IBMQ device and their calibration data}

For each device listed in~\cref{tab:DataDates}, we provide below the topology map showing its qubit connection, calibration data taken around the time the computation data was collected, the range and median of its readout assignment error and CNOT error, and the qubit pairs investigated in this work.

\begin{table*}
    \centering
    \begin{tabular}{|c|c|c|c|c|}
    \hline
    \multirow{2}{*}{Device} & \multicolumn{2}{|c|}{Readout error} & \multicolumn{2}{|c|}{CNOT error} \\
    \cline{2-5}
    & Median & Range & Median & Range \\
    \hline
    Washington & $1.290\times10^{-2}$ & $1.900\times10^{-3}$ to $4.854\times 10^{-1}$ & $1.234\times10^{-2}$ & $5.999\times10^{-3}$ to $1.000$ \\
    \hline
    Geneva & $2.930\times10^{-2}$ & $7.300\times10^{-3}$ to $3.683\times10^{-1}$ & $5.457\times10^{-2}$ & $3.872\times10^{-3}$ to $1.000$ \\
    \hline
    Cairo & $1.190\times10^{-2}$ & $6.000\times10^{-3}$ to $1.221\times 10^{-1}$ & $9.815\times10^{-3}$ & $4.436\times10^{-3}$ to $1.000$ \\
    \hline
    Hanoi & $1.150\times10^{-2}$ & $5.800\times10^{-3}$ to $8.690\times 10^{-2}$ & $7.465\times10^{-3}$ & $2.982\times10^{-3}$ to $1.000$ \\
    \hline
    Mumbai & $1.860\times10^{-2}$ & $1.150\times10^{-2}$ to $1.115\times 10^{-1}$ & $7.804\times10^{-3}$ & $4.201\times10^{-3}$ to $1.888\times10^{-2}$ \\
    \hline
    \end{tabular}
    \caption{Median and range of the readout assignment error and CNOT error for each IBMQ device.}
    \label{tab:errorSummary}
\end{table*}

\newpage
\section{Other miscellaneous results}\label{App:MiscResults}

We give here the results analogous to \cref{tab:smallPval}, but with a more stringent (smaller) significance level.

\begin{table}[h]
    \centering
    \begin{tabular}{|c|c|c|c|c|c|c|}
    \hline
    Device & Qubits[Circuit] & $\Q^3$ & $\NS$ & $\NSnotAtoB$ & $\NSnotBtoA$ & $\L$    \\
    \hline
    \multirow{3}{*}{Washington} &
      12,17 [$\CirNL$] & 0 & 0 & 0 & 1 & 0 \\
    &  38,39 [$\CirNL$] & 0 & 0 & 0 & 1 & 0 \\
    &  91,98 [$\CirNL$] & 1 & 1 & 0 & 0 & 1 \\
    \hline
    \multirow{2}{*}{Geneva} &
    14, 16 [$\CirNL$] & 0 & 0 & 0 & 1 & 0 \\
    &  21, 23 [$\CirNL$] & 10 & 10 & 1 & 17 & 8 \\
    \hline
     Cairo & 13,14 [$\CirNL$] & 1 & 1 & 1 & 0 & 100 \\
    \hline
    Hanoi &  5,8 [$\CirNL$]   & 1 & 1 & 0 & 0 & 41 \\
    \hline
    \end{tabular}
    \caption{Summary of results parallel to those presented in \cref{tab:smallPval} but with the significance level set at the more stringent value of $\alpha = 0.01$.}
    \label{tab:smallPval001}
\end{table}

\section{Results for a product-state generating circuit}\label{App:Local}

\begin{figure}[h]
    \centering    \includegraphics[width=0.45\columnwidth]{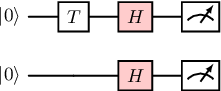}\\
\caption{Quantum circuits generating a non-Bell-inequality-violating correlation. $T = \mathrm{diag}(1, e^{i\pi/4})$ refers to the $\pi/8$-phase gate. The symbols and shading carry the same meaning as those in \cref{Eq:CNL}.}
\label{Eq:CL}
\end{figure}

The circuits $\CirL$ for generating a Bell-local correlation are given in \cref{Eq:CL}. The ideal correlation resulting from this circuit is that obtained by measuring Pauli-$Z$ and Pauli-$X$ on the state $\ket{00}$. Note that while the $T$ gate is irrelevant in theory, the fact that it is performed in the circuit can still have a nontrivial consequence in the experiment.

\begin{table}[h]
    \centering
    \begin{tabular}{|c|c|c|c|c|c|c|}
    \hline
    Device & Qubits [Circuit] & $\Q^3$ & $\NS$ & $\NSnotAtoB$ & $\NSnotBtoA$ & $\L$    \\
    \hline
     \multirow{4}{*}{Washington} &
        12,17 [$\CirL$] & 0 & 0 & 1 & 0 & 0 \\
     &  38,39 [$\CirL$] & 1 & 1 & 2 & 0 & 1 \\
     &  79,91 [$\CirL$] & 1 & 1 & 0 & 2 & 1 \\
     &  91,98 [$\CirL$] & 2 & 2 & 0 & 2 & 1 \\
    \hline
      \multirow{2}{*}{Cairo} &
       13,14 [$\CirL$] & 0 & 0 & 2 & 0 & 0 \\
     & 23,24 [$\CirL$] & 1 & 1 & 1 & 1 & 1 \\
	\hline
     \multirow{4}{*}{Hanoi} &
        5,8 [$\CirL$]    & 2 & 2 & 0 & 1 & 1 \\
     &  6,7 [$\CirL$]    & 1 & 1 & 0 & 0 & 0 \\
     &  11,14 [$\CirL$]  & 2 & 2 & 2 & 0 & 0 \\
     &  19,20 [$\CirL$] & 0 & 0 & 0 & 1 & 0 \\
    \hline
    \end{tabular}
    \caption{Summary of results analogous to those presented in \cref{tab:smallPval} but with the circuits considered being those given in \cref{Eq:CL}.}
    \label{tab:smallPvalL}
\end{table}

\begin{table}[t]
    \centering
    \begin{tabular}{|c|c|c|c|c|c|c|c|}
    \hline
    Device & Qubits [Circuit] & $\Q^3$/ $\NS$  & $\NSnotAtoB$ & $\NSnotBtoA$  & $\L$   \\
    \hline
    \multirow{4}{*}{Washington}
    &  12,17 [$\CirL$] & -  & 49 & - & -  \\
    &  38,39 [$\CirL$] & 98  & 57, 98 & - & 98  \\
    &  79,91 [$\CirL$] & 78  & - & 30, 78 & 78  \\
    &  91,98 [$\CirL$] & 38, 64  & - & 38, 64 & 38  \\
    \hline
      \multirow{2}{*}{Cairo}
     & 13,14 [$\CirL$] & -  & 64, 76 & - & -  \\
     & 23,24 [$\CirL$] & 30  & 59 & 77 & 30  \\
    \hline
    \multirow{4}{*}{Hanoi}
    &  5,8 [$\CirL$]   & 11, 61  & - & 11 & 61  \\
    &  6,7 [$\CirL$]   & 58  & - & - & -   \\
    &  11,14 [$\CirL$]  & 27, 68  & 27, 68 & - & 27  \\
    &  19,20 [$\CirL$] & - & - & 96 & -  \\
    \hline
    \end{tabular}
    \caption{Further details about the instances of Bell tests giving the results reported in \cref{tab:smallPvalL}. Under the third column to the rightmost, we list the Bell test number implemented on the respective device and qubit pair that shows a violation of the corresponding null hypothesis. To simplify the presentation, we have put the identical results for $\Q^3$ and $\NS$ under the same column.
}
\label{tab:smallPvalL:details}
\end{table}

\end{CJK*}

\end{document}